\documentclass{sig-alternate}
\usepackage{url}
\usepackage{graphicx}
\usepackage{epstopdf} 
\usepackage{algorithmic}
\usepackage{epsfig}
\usepackage{mdwlist}
\usepackage{flushend}

\DeclareMathOperator*{\argmax}{argmax}

\begin{document}
\title{Iterative Expectation for Multi Period Information Retrieval}
 \numberofauthors{1} \author{
  \alignauthor
  Marc Sloan  and Jun Wang\\
  \affaddr{Department of Computer Science}\\
  \affaddr{University College London}\\
  \email{\{M.Sloan, J.Wang\}@cs.ucl.ac.uk} \alignauthor }
\maketitle

\begin{abstract} 
Many Information Retrieval (IR) models make use of offline statistical techniques to score documents for ranking over a single period, rather than use an online, dynamic system that is responsive to users over time. In this paper, we explicitly formulate a general Multi Period Information Retrieval problem, where we consider retrieval as a stochastic yet controllable process. The ranking action during the process continuously controls the retrieval system's dynamics, and an optimal ranking policy is found in order to maximise the overall users' satisfaction over the multiple periods as much as possible.  Our derivations show interesting properties about how the posterior probability of the documents relevancy evolves from users feedbacks through clicks, and provides a plug-in framework for incorporating different click models. Based on the Multi-Armed Bandit theory, we propose a simple implementation of our framework using a dynamic ranking rule that takes rank bias and exploration of documents into account. We use TREC data to learn a suitable exploration parameter for our model, and then analyse its performance and a number of variants using a search log data set; the experiments suggest an ability to explore document relevance dynamically over time using user feedback in a way that can handle rank bias.
\end{abstract}

\section{Introduction}
In Information Retrieval (IR) research, we study mathematical models of IR systems because they provide formal and quantitative tools for us to understand the underlying retrieval mechanisms, and at the same time, lead to the development of practical retrieval algorithms and systems. Mainstream IR theories have been largely devoted to the offline evaluation of corpora to estimate document relevancy, such as the statistical language models and relevance models that score a query by analyzing term statistics \cite{LM_Zhai2008} or link analysis approaches that look at the long term stabilised visit rates of web documents \cite{Franceschet:2011}.

Alternatively, we propose an online approach that is able to dynamically learn, over multiple periods, the correct ranking of documents for a query with or without any preceding evaluation. This variation of the IR problem can be applied in situations where document content is unclear or incomplete for prior analysis (for instance in collaborative filtering); where new documents of unknown relevance are routinely added (such as information filtering); or simply as a complement to existing data and techniques (web search). We formally express \textit{Multi Period Information Retrieval} as:

\vspace{5pt}

\emph{``An information retrieval system receives a query over time $t=1\ldots T$, where $t$ denotes a single period when the query is shown to a user. For each period the system returns a set of ranked documents and observes user feedback in the form of clickthroughs. How do we determine an optimal ranking policy sequentially that maximises the overall user's satisfaction for that query over the multiple periods of time?''}

\vspace{5pt}

Our study in this paper is a theoretical one; we offer a general solution framework by using an \textit{optimal control} formulation
\cite{athans2006optimal}. Any type of dynamically controlled system requires a control signal \cite{control-book}, which we regard as the rank action at time $t$, where its state, in our case, denotes the system's belief about the documents' relevancy. Some form of sensor (such as user feedback) is also required so that the system can adjust its state (thus its belief about the relevancy) to its changing environment. The rank action at a given time alters the system states by considering the system's dynamics (in this case the time evolution of the belief about the documents relevancy). As a result, we find that the optimisation framework is flexible and many click models, such as \cite{clickmodels,dcm}, can be naturally integrated. It is worth noting that our theoretical framework is different with regard to the online learning approaches in \cite{joachims:online-learning,Yue:2009}, where the goal is to optimize over pairwise preference learning, whilst ours is to estimate the probability of relevance over time, making it more closely related to the classic information retrieval methodologies such as relevance models \cite{Robertson:1988,Lavrenko:2001} and relevance feedback \cite{Rocchio}.

Under this framework, three critical issues are addressed. Firstly, the system's state is represented as the posterior probability of a
document being relevant, and we propose an iterative update mechanism to update the posterior probability about the documents relevancy from users' feedbacks.  The derivation gives insights into how rank bias influences the evolution of the probability of relevance. 
Secondly, the system's dynamic function is derived, which shows how the belief (thus the posterior probability) evolves according to the rank decision made in the past and the user feedback received so far.
Lastly, equipped with the dynamic function, we demonstrate its use by developing simple ranking rules which incorporate the recent results from multi-armed bandit machine research \cite{auer}. The dynamic function is flexible enough to incorporate parameters from a number of click models, and the degree of exploration used by the ranking rules can be adjusted.

We conduct two types of experiment. Firstly, experiments incorporating simulated users based on TREC relevance judgement data allow us to tune the algorithm over a range of parameters and provides some preliminary analysis on the effect of exploration on the performance of various click models. We follow this experiment by using the Yandex Relevance Prediction Challenge data set to test our algorithm in a realistic, practical setting using search log data in order to demonstrate its capability in learning relevance rankings. We also illustrate the flexibility of the model by testing a number of variants, for instance, a version that incorporates new documents dynamically and another that learns from click log data in advance. 

This paper is organised as follows: we start with the related work in Section~\ref{relatedWork}, multi period IR
modeling will be presented in Section~\ref{formulation}, Section~\ref{experiments} will discuss and display the experimental results
and the evaluation and conclusions will be in Section~\ref{conclusions}.

\section{Related Work}
 \label{relatedWork}

 The multi period IR problem is a general one and completely solving it requires various aspects of research. This section provides a short overview of the related work.

The concept behind multi period information retrieval, that of dynamically using user feedback to improve a search ranking over time, is built upon the ideas behind relevance feedback where the aim is typically to extract information from the user and use it to update the IR system. Much of the literature concerns the various methods of acquiring this information and fall broadly into two categories:
\begin{description}
\item [Explicit] Where the user directly rates a document or a search ranking for its relevance (often to the detriment of the user experience). The Rocchio algorithm is commonly used with this form of feedback \cite{Rocchio}.
\item [Implicit] User actions, such as time spent on search page, clickthroughs and actions following a clickthrough \cite{Dupret:2010:MEI:1718487.1718510}, are recorded whilst users are in the process of satisfying their information need. Whilst it is important that indirect observations of user behaviour must be interpreted correctly to infer feedback, they have been shown to correlate with explicit relevance judgements \cite{Fox:Implicit_measures}.
\end{description}
For our model we chose clickthroughs as our method of implicit user feedback as it is a measure that is abundant, cheap, readily available and there has been a plethora or research into how to interpret them. 

\textit{Clickthroughs} have established themselves as the most common way to infer user satisfaction for a given IR task (for instance, web search) \cite{Fox:Implicit_measures, Agichtein:2006:IWS}; nonetheless, they are not a perfect representation of user interest and document relevance and are subject to noise and bias \cite{Falk:clicks_no_good}. It has been shown that aside from the relevance of the displayed document, the largest influence on whether a user clicked on a document is its rank \cite{Joachims/etal/05a}. Various \textit{click models} have been proposed, from the basic mixture model, examination hypothesis and cascade models \cite{clickmodels} (explored in this paper, along with the dependent click model \cite{dcm}) to the recent General Click model \cite{Zhu:GCM}; these models being Bayesian representations of the probabilities associated with a users typical behaviour. What these models have in common is that they attempt to disambiguate what a clickthrough is actually inferring in the context of a set of search results, and as such a click model is essential in any dynamic system employing clickthroughs as implicit feedback, as evidenced on learning click models in \cite{Agichtein:2006:LUI}. 

The key problem in this paper is how to devise optimal ranking strategies over time which dynamically improves a ranking algorithm and in the end maximises a certain expected utility. One of the relevant formulations is the \textit{Multi-Armed Bandit Problem}, which is a classic statistical resource allocation problem. The usual analogy is that of a casino with multiple one-armed bandit slot machines, each with a different probability distribution of rewards, and the problem is to find the best strategy for exploring the distributions of the bandits, whilst also exploiting the highest paying bandit. Building on previous work on non-optimal but asymptotic solutions, \cite{auer} designed practical, linear-time algorithms including the popular UCB (Upper Confidence Bound) algorithm which much of this work is based on. Another important variant is the Exp3 algorithm \cite{exp3}, which is designed to provide asymptotic performance in the worst case scenario of an adversary setting the bandit rewards (rather than stochastic or deterministic rewards).

The UCB algorithm is an ongoing area of research and has improved variations, offering lower regret bounds (a typical evaluation metric) and better optimal performance \cite{ImprovedUCB, regretBounds}. Also, multi-slot Multi-Armed Bandit (MAB) are another active area \cite{cKaleRS10}, with a particular focus on stochastic algorithms in the adversarial setting. MAB's are increasingly being used in IR, such as using one to display web adverts or news stories \cite{Chakrabarti:mortal}, and also \cite{Radlinski} where the authors used it to provide a diversified rank solution. 

A related theoretical development can be found in \cite{DBLP:journals/ir/Fuhr08}, which extends the PRP to cover interactive retrieval, the goal being to design interactive strategies within a single search session for a given user. A similar idea to this was proposed by \cite{Brandt:2011}, providing an interactive method of displaying diverse results whilst retaining high recall. These two methods interpret \textit{dynamic} IR as an interactive, single user experience, whereas we are instead concerned with the dynamics of the IR system itself over time with a population of users. The term can also refer to the dynamic nature of document content (such as for news websites), which \cite{temporal-dynamics} exploits in order to improve relevancy rankings.

Other work that focuses on using clickthroughs in an iterative fashion to learn document rankings include \cite{xue:optimize-using-clicks}'s work in using clickthroughs in search logs to infer the similarity between pairs of queries and pairs of documents. More recently, \cite{joachims:online-learning,Yue:2009} also propose an alternative online algorithm for learning document rankings via user feedback, a key difference being their focus on preference learning (pairwise comparison) as opposed to our focus on probability of relevance and rank bias, which gives our algorithms flexibility with click models. In this regard, our statistical framework can be considered as the multi period extension of the classic relevance retrieval methodologies proposed in  \cite{Robertson:1988,Lavrenko:2001}.

\section{Multi Period IR Modeling} \label{formulation} 

\subsection{An Optimal Control Formulation}
We investigate a scenario for a given information need (such as a query) where there are \(N\) documents that need to be ranked. For each time period $t$ (or for each search of the given query), a fixed number \(M\) documents (where \(0 < M \leq N\)) are retrieved and displayed to the user. The rank decision (or \textit{rank action}) is denoted as vector
\[
\begin{gathered}
\mathbf{d}(t)=\{d_1(t),\ldots,d_M(t)\} \in\{1,\ldots,N\}^{M},
\end{gathered} 
\]
where \(d_i(t)\) is the document retrieved at rank \(i\) and at time~\(t\) (where $d_i(t) \neq d_j(t) \ \forall\  i \neq j$).

We view the IR problem as a dynamic and controllable process, in which the rank action $\mathbf{d}$ acts as our input into the system, and we record clickthroughs $R(\mathbf{d}(t))$ as the output. Given that we cannot directly observe the probability of relevance for each document, we can instead observe the clickthroughs at each rank, which acts as the systems output signal, and define $R(\mathbf{d}(t))$ as the number of clickthroughs observed for ranked documents $\mathbf{d}(t)$. 

The goal of multi period IR is to find the optimal rank action $\mathbf{d^*}$ that maximises the users satisfaction over time, e.g.,  the expected number of clickthroughs (biased by rank) over time, which is given by
\begin{align}
 \nonumber \mathbf{d^*} = (\mathbf{d}^*(1),\ldots,\mathbf{d}^*(T))  
&= \argmax _{\mathbf{d}(1)\ldots\mathbf{d}(T) }E\left[ \sum\limits_{t = 1}^T {R(\mathbf{d}(t))}\right] \\
&= \argmax _{\mathbf{d}(1)\ldots\mathbf{d}(T) } \sum\limits_{t = 1}^T E[{R(\mathbf{d}(t))}]  \label{objFunction}
\end{align}
 Eq.~(\ref{objFunction}) is a result of assuming that the clickthroughs are independent and identically distributed (i.i.d.), which is reasonable as for each $t$ the system might receive independent clicks (note that the clicks over a ranked list may not be independent) from a random user. It should also be noted that the expectation is over the probability of click and the observed clicks random variable, which will be explored fully in the next section where we will describe an algorithm to iteratively maximise this expectation.

\subsection{Iterative Expectation} 
\label{rankedBandit}
We are now concerned with maximising the expected number of clicks \(E\left[R (\mathbf{d}(t))\right]\), which is achieved when using a rank action that displays in decreasing order those documents that have the highest probability of relevance (as defined in the Probability Ranking Principle \cite{JD:1977:Robertson:PRP}). We can estimate the expected number of clicks for a given ranking by using a \textit{click model}. 

For a ranking $\mathbf{d}$ we have observations over time about the clickthroughs in different rank positions $\{C_i(t)\}_{t=1}^{T}$, where $C \in \{0, 1\}$ is a binary variable representing a click. These can be used to infer the expected clicks over time:
 \begin{align}\label{clickModelEq}
   \nonumber  \sum_{t=1}^T E[R(\mathbf{d}(t))] =&\sum_{t=1}^T\sum_{i=1}^{M}\sum_{C_i(t)}C_i(t) p(C_i(t)|d_i(t))\\
   =&\sum_{t=1}^T\sum_{i=1}^{M} p(C_i(t)=1|d_{i}(t))\end{align} 
However, we do not know whether a user clicking on a document was due to
 the fact that it is relevant or simply because it has a high ranking
 position. Mathematically, we assume that the click is generated by a
 mixture of two binomial distributions, where a hidden binary variable
 is used to represent the membership, i.e. $S_i=0$ denotes it is due to
 the rank bias at rank $i$, and $S_i=1$ due to the document relevance.  This results in
 the following conditional probability of the click (the time variable $t$ is implied when omitted):
\begin{align}\label{IE-1}
 p(C_i|d_{i})  &= p(C_{i}|d_{i},S_i =1)p(S_i=1) \\
\nonumber &\ \ \ \ \ \ \ \ \ + p(C_{i}|S_i = 0)p(S_i = 0) \hfill \\
\nonumber &= p(C_{i}|d_{i})p(S_i = 1) + p(C_{i}|S_i=0)p(S_i=0) \\
\nonumber &= r_{d_i}^{C_{i}}(1-r_{d_i})^{1-C_{i}}\pi_i + b_{i}^{C_{i}}(1-b_{i})^{1-C_{i}}(1-\pi_i)
 \end{align} 
Here, three parameters have been defined:
\begin{align}
  r_{d_i} \equiv p(C_i=1|d_i ), b_i \equiv p(C_i=1|S_i =0), \pi_i \equiv p(S_i = 1)  \label{parameters} \end{align}
where $r_{d_i}$ is the probability of relevance for document $d_i$, $b_i$ is the \textit{bias} associated with rank $i$ and $\pi_i$ is the probability that a user will click due to the document's relevance rather than blindly due to the rank i.e. the \textit{trust} the user has in the search engine.
A mixture click model has been similarly defined and
evaluated against real click-through data in \cite{clickmodels}, and it
should be emphasized that the mixture model presented here is a more
general one. A heuristic background click model in \cite{Agichtein:2006:LUI} is in fact a case where $\pi_i=1$ and $b_i$ is considered as the background click rate, and it can be shown that other click models such as the Examination Hypothesis model \cite{clickmodels} and Dependent Click model \cite{dcm} are also special cases in the formulation by setting different types of parameters in Eq.~(\ref{parameters}).

We have the following EM algorithm to estimate the parameters up to time $T$
\cite{em-tutorial}:\\
\textbf{E Step}:
\begin{align}
p(S_i|C_i) = \frac{{p(C_i|S_i )p(S_i )}}
{p(C_i |S_i=1 )p(S_i =1)+p(C_i |S_i=0 )p(S_i =0)}
\end{align}
\textbf{M Step}:
\begin{align}
\hat{r}_{d_i} &= p(C_i(T)=1|d_{i} ) = \frac{{\sum\limits_{t=1}^{T} {C_i(t)p(S_i  = 1|C_i(t))} }}
{{\sum\limits_{t=1}^{T} {p(S_i = 1|C_i(t))} }} \label{M-1}\\
b_i &= p(C_i(T)=1|S_i  = 0) = \frac{{\sum\limits_{t=1}^{T} {C_i(t)p(S_i  = 0|C_i(t))} }}
{{\sum\limits_{t=1}^{T} {p(S_i  = 0|C_i(t))} }}\label{M-2}\\
\pi_i &= p(S_i=1) = \frac{1}
{T}\sum\limits_{t=1}^{T} {p(S_i=1 |C_i(t))} \label{M-3}
\end{align}

where the E step is obtained by applying Bayes' Rule. The M step can be obtained by maximizing the lower bound of the following likelihood function: 
\begin{align*}
  \nonumber & L(\{r_{d_i}\},\{\pi_i\},\{b_i\}) =\prod_{t=1}^T p(C_i(t)|d_i)\propto  \sum_{t=1}^T \log p(C_i(t)|d_i)\\
 \nonumber &= \sum_{t=1}^T \log [ r_{d_i}^{C_i(t)}(1-r_{d_i})^{1-C_i(t)}\pi_i\\
\nonumber &\ \ \ \ + b_{i}^{C_i(t)}(1-b_{i})^{1 -C_i(t)}(1-\pi_i)]\\
  \nonumber &\geq \sum_{t=1}^T \Big(p(S_i=1|C_i(t))\log \frac{ r_{d_i}^{C_i(t)}(1-r_{d_i})^{1-C_i(t)}\pi_i}{p(S_i=1|C_i(t))}\\
  \nonumber &\ \ \ \ + p(S_i=0|C_i(t))\log \frac{ b_{i}^{C_i(t)}(1-b_{i})^{1-C_i(t)}(1-\pi_i)}{p(S_i=0|C_i(t))}\Big)\\
  \nonumber& \propto  \sum_{t=1}^T \Big(p(S_i=1|C_i(t))\log (r_{d_i})^{C_i(t)}(1-r_{d_i})^{1-C_i(t)}\pi_i\\
  \nonumber &\ \ \ \ + p(S_i=0|C_i(t))\log (b_{i})^{C_i(t)}(1-b_{i})^{1-C_i(t)}(1-\pi_i)\Big)
\end{align*}
where the lower bound holds due to Jensen's inequality \cite{em-tutorial}. Maximizing the last step with respect to the three parameters respectively obtains the M step.  

This algorithm requires iterative steps for each time $t$ and we can simplify the update procedure for Eq.~(\ref{M-1}) by fixing the E step and click model parameters $\pi_i$ and $b_i$, and introducing the following two variables:
\begin{align}
\nonumber &\alpha_i  \equiv p(S_i=1|C_i=1) \\
\nonumber =& \frac{{p(C_i=1|S_i=1 )p(S_i=1 )}}
{p(C_i= 1 |S_i=1 )p(S_i =1)+p(C_i =1 |S_i=0 )p(S_i =0)}\\
=&\frac{\hat{r}_{d_i} \pi_i}{\hat{r}_{d_i}\pi_i+b_{i} (1- \pi_i)}\\\nonumber\\
\nonumber &\beta_i  \equiv  p(S_i=1|C_i=0) \\
\nonumber =& \frac{{p(C_i=0|S_i=1 )p(S_i=1 )}}
{p(C_i=0 |S_i=1 )p(S_i =1)+p(C_i =0 |S_i=0 )p(S_i =0)}\\
=&\frac{(1-\hat{r}_{d_i})\pi_i}{(1-\hat{r}_{d_i})\pi_i+(1-b_{i} )(1- \pi_i)}
 \end{align}
 where $\hat{r}_{d}$ is the probability of click for document $d$
 estimated from past observations. 

The probability of click Eq.~(\ref{M-1}) is subsequently updated by:
\begin{align}
 \hat{r}_{d_i}(T)= p(C_i(T)=1|d_i(T) ) = \frac{{\sum\limits_{t=1}^{T} {  C_i (t)\alpha_i(t)^{C_i (t)}\beta_i(t)^{1-C_i(t)} } }}
{\sum\limits_{t=1}^{T} {  \alpha_i(t)^{C_i(t)}\beta_i(t)^{1-C_i(t) }
     }}
\end{align}
Which can be iteratively obtained over time, giving us
\begin{align}\label{IE}
\hat{r}_{d_i}(T) =& \hat{r}_{d_i}(T - 1) \frac{\gamma_{d_i}(T-1)}{\gamma_{d_i}(T)}+ C_i(T)\left(1-\frac{\gamma_{d_i}(T-1)}{\gamma_{d_i}(T)}\right)\hfill
\end{align} 
where \(\gamma_{d_i}(t)\equiv\sum\limits_{k = 1}^{t} { \alpha_i(k)^{C_i(k)}\beta_i(k)^{1-C_i(k) }}\)
serves as an ``effective'' number of impressions, balancing the influence between the rank bias and probability of relevance.

\textbf{Analysis: }$\alpha_i$ and $\beta_i$ are considered as
 ``effective counts'' and differentiate rank positions when receiving
 a click and non-click, respectively.  $\alpha_i$ is an increasing
 function of $i$ and is larger if receiving a click on a lower ranked document, rewarding the fact that a user makes
 efforts to reach that rank because it is unlikely that the click
 is due to the rank bias. By contrast, if a document is clicked in the top rank, the effective
 count is rather small as the click might just be due to the rank
 bias, so by the same token we also notice that $\beta_i$ is a
 decreasing function with respect to $i$ i.e. for a higher ranked
 document, $i$ is small and thus $\beta_i$ is large. If such a high
 ranking document is not clicked, we effectively penalize the document
 more by adding large $\beta_i$ in the denominator, and as the rank
 position increases, the penalty becomes small as well.  In summary, a
 non-click on the document at high rank or a click on a low ranked
 document is an important observation and should give a large
 effective count when updating our belief about the probability of a
 click.

Eq.~(\ref{IE}) provides the solution to our multi period IR problem, it is a dynamical function of the probability of relevance, the dynamics of
which are relevant to the control signal ${\mathbf{d}}$, which is dependent on the previous observations of clickthroughs.  
Adjusting parameters $\pi_i$ and $b_i$ also allows us to incorporate some different click models; to illustrate the flexibility of the proposed 
framework, in our experiments we provide alternative schemes by plugging
in the Examination Hypothesis and Dependent Click models introduced in \cite{clickmodels, dcm}.

\subsection{Ranking Rule Over Time}
Based on the previous section, our multi period information retrieval problem can be restated as follows:
\begin{align}
\nonumber  \mathbf{d}^*&= \argmax _{\mathbf{d}(1)\ldots\mathbf{d}(T) )} \sum\limits_{t = 1}^T {\sum_{i=1}^M p(C_i=1|d_{i}) }  \\
  &= \argmax _{\mathbf{d}(1)\ldots\mathbf{d}(T) } \sum\limits_{t = 1}^T {\sum_{i=1}^M  \left\{\hat r_{d_i} \pi_i + b_i (1-\pi_i) \right \}} \hfill 
\end{align}
subject to Eq.~(\ref{IE}).

The optimisation is subject to the constrained dynamic function for $\hat r_{d} $. Ideally we wish to seek a dynamic programming solution \cite{athans2006optimal} to find the optimal control rule by modeling the problem as a Partially Observable Markov Decision Process, but this will be the subject of future research. Instead, an index solution (assigning index scores to documents individually and then rank them according to the scores) is proposed to demonstrate the use of the multi period IR formulation by following a multi-armed bandit machine approach.

\subsubsection{Iterative Expectation (UCB-IE)  Algorithm }
The following algorithm is a variant of the popular UCB1 algorithm \cite{auer}, where we attempt to optimise over the values for $\hat{r}_{d}$, which is updated using Eq.~(\ref{IE}). We've chosen to proceed with the UCB algorithm due to its simplicity, familiarity and demonstrative value for our framework, and leave it to future research to incorporate more complex online learning algorithms. 

The UCB-IE algorithm:

\begin{enumerate}
\item Set all $\gamma_d(t) = 1$ (to avoid division by zero) and all $\hat{r}_{d}$ to some prior value between 0 and 1, these priors could be learned relevancy probabilities or simply a flat prior of 0.5, either way the outcome will be the same, all that will be affected is the learning speed. Also, set parameters $\pi_i$ and $b_i$ to that of your click model, and time step $t = 1$
\item For all documents $d$, calculate
\begin{align}
\Lambda_d = \hat{r}_d + \lambda\times\sqrt{\frac{2\ln{t}}{\gamma_d(t)}}\label{gammaEq}
\end{align}
where $\lambda$ is a constant indicating the preferred degree of exploration. 
\item Set $\mathbf{d}(t)$ to be the $M$ documents with the highest $\Lambda_d$ values in decreasing order and display them to the user
\item Record clicks $C_i(t)$ and update variables accordingly, for $i = 1\to M$
\begin{align*}
\alpha_i(t) &=\frac{\hat{r}_{d_i}(t - 1) \pi_i}{\hat{r}_{d_i}(t - 1)\pi_i+b_i (1- \pi_i)} \\
\beta_i(t) &=\frac{(1-\hat{r}_{d_i}(t - 1))\pi_i}{(1-\hat{r}_{d}(t - 1))\pi_i+(1-b_i )(1- \pi_i)}\\
\gamma_{d_i}(t) &= \gamma_{d_i}(t - 1) + \alpha_i(t)^{C_i(t)}\beta_i(t)^{1-C_i (t) } \\
\hat{r}_{d_i}(t )&= \hat{r}_{d_i}(t - 1) \frac{\gamma_{d_i}(t - 1)}{\gamma_{d_i}(t)}+ C_i(t)\left(1-\frac{\gamma_{d_i}(t - 1)}{\gamma_{d_i}(t)}\right)
\end{align*}
\item Repeat steps 2 to 5 for $t = 2\ldots T$
\end{enumerate}

The Mixed Clicks, Examination Hypothesis  and Dependent Click models can be incorporated into the above algorithm by setting the parameters $\pi_i$ and $b_i$ accordingly:
\begin{description}
\item [Mixed Clicks] $\pi_i , \text{   }b_i$
\item [Examination Hypothesis] $\pi_i \equiv \eta_{i},  \text{   }b_i \equiv 0$
\item [Dependent Click Model] $\pi_i \equiv \prod_{j = 1}^{i - 1} ( 1 - \hat{r}_{d_j} + \eta_j \hat{r}_{d_j}),  {b_i \equiv 0}$
\end{description}
where $\eta$ is a click model specific rank bias parameter.

Using this algorithm, we attempt to learn the optimal $\hat{r}_{d}$, representing the documents with the highest probability of relevance. Due to the rank bias, we adjust the reward given by a click to reflect the likelihood of the click given its position, and in a similar way use the effective count $\gamma_d(t)$ rather than the actual number of impressions (such as in UCB1). Thus, documents that are displayed at lower ranks will have less `impressions' than higher ranked documents, and so will still encourage exploration of these documents into the higher ranks. 

Exploration is encouraged by supplementing the probability of relevance in Eq.~(\ref{gammaEq}) with an index term. This term grows larger with each time step $t$, or smaller with increasing $\gamma$ i.e. the term will be small for documents that are frequently displayed, and will gradually grow larger for those that are not, until it is large enough to cause the document to be displayed. The exploration parameter $\lambda$ can be tuned to optimize the amount of exploration that occurs when generating a document ranking, including being set to 0, causing the UCB-IE algorithm to act myopically and greedily rank documents in strictly decreasing order of relevancy. 

\section{Experiments}
\label{experiments}

In the previous section we studied the theoretical properties of document ranking over time under the assumption of certain click models. This section continues the study by evaluating the resulting practical ranking strategies by first considering a simulation and then using click log data. We primarily intend to 1) show convergence to an optimal ranking that can maximise evaluation metrics such as MAP and nDCG over time; 2) determine the benefit of exploration by contrasting UCB-IE with a myopic variant, and subsequently find an ideal value for $\lambda$; and 3) demonstrate how the model can be used in conjunction with typical click log data to provide good search ranking performance. 

Three resulting ranking strategies were evaluated. Namely, we have the UCB algorithm with Interactive Expectation using the Mixed-Click (denoted as UCB-IE-MC), Examination Hypothesis (UCB-IE-EH) and Dependent Click (UCB-IE-DCM) models. We chose these models as we believe they reflect the reality well as they have been thoroughly studied using real clickthrough data (e.g. \cite{clickmodels}) and they could be plugged into our framework.

The ideal evaluation of an online learning algorithm that incorporates rank bias would require access to an operational search engine to experiment on. With no access to one, we have instead performed two evaluations, the first using simulated users and TREC relevance judgement data, the second using Yandex click log data, both of which are explored further in the following subsections. 

\begin{table*}[ht]
 \centering
\begin{tabular}{c | c | c | c | c | c | c | c | c}
 \multicolumn{9}{c}{\textbf{MAP}} \\
   \hline  \hline
$\lambda$ & 0 & 0.001 & 0.01 & 0.05 & 0.1 & 0.2 & 0.5 & 1 \\
\hline
\textbf{MC} & 0.8330 & 0.8344 & 0.8388 & 0.8428 & 0.8586 & \textbf{0.8849} & 0.4888 & 0.1112   \\
\textbf{EH} & 0.7979 & 0.8024 & 0.8083 & 0.8391 & \textbf{0.8625} & 0.8612 & 0.1875 & 0.0930 \\
\textbf{DCM} & 0.8110 & 0.8022 & 0.8118 & 0.8436 & \textbf{0.8604} & 0.8576 & 0.8274 & 0.6707 \\
\hline 
 \multicolumn{9}{c}{\textbf{nDCG@10}} \\
\hline \hline
\textbf{MC} & \textbf{0.9982} & 0.9977 & 0.9977 & 0.9965 & 0.9914 & 0.9771 & 0.6141 & 0.2364 \\
\textbf{EH} & 0.9940 & \textbf{0.9965} & 0.9942 & 0.9931 & 0.9793 & 0.9331 & 0.3391 & 0.2068 \\
\textbf{DCM} & \textbf{0.9999} & \textbf{0.9999} & \textbf{0.9999} &	0.9997 & 0.9993 & 	0.9868 & 0.9464 & 	0.8056
\end{tabular}
\caption{Summary of mean final MAP and nDCG@10 values after $T = 500$ time steps for each click model \textbf{UCB-IE} variant, and for each value of the exploration parameter $\lambda$. Maximum values are in bold. }
\label{table1:TREC}
\end{table*}

\subsection{TREC Simulation Analysis}

We first evaluated whether our model was capable of learning a correct ranking of documents over time, and also the effect of exploration on finding an optimal ranking. We developed an experiment using TREC relevance judgement data and simulated stochastic users who clicked according to click models that we set. This allowed us to perform the experiment in a controlled setting (free of the assumptions that restricted our click log data experiment) and optimize and analyse the exploration parameter $\lambda$.

To simulate a realistic collection of documents and associated relevance values we used `TREC 2001 Web Ad Hoc qrels' relevance judgements from the TREC-10 Web Ad Hoc Retrieval Track. We ran the algorithm over 50 different topics (representing 50 queries), each of which contained an average of 1408 documents and 41.53 documents judged relevant (the judged relevance being unknown to the algorithms). Judgements were graded either 0, 1 or 2, which were normalised to give probability of relevance values 0, $\frac{1}{2}$ and 1. 

At each time step a ranking of $M = 10$ documents was generated by each of the UCB click model variants, and a simulated user then examined and clicked on the documents in the ranking according to the corresponding click model. With no data to learn the click model parameters from, we set the following click model parameters (the value of 0.8 was picked as it delivered consistent performance across all click models and topics): 
\begin{description} 
\item [Mixed Click Model] $\pi_i \equiv 0.8$, $b_i \equiv 0.8^{i-1}$
\item [Examination Hypothesis] $\pi_i \equiv 0.8^{i - 1}$, $b_i \equiv 0$ 
\item [Dependent Click Model] $\pi_i \equiv \prod_{j = 1}^{i - 1} ( 1 - \hat{r}_{d_j} + 0.8_j \hat{r}_{d_j})$, \\${ b_i \equiv 0}$ (updated at each time step)
\end{description}

For each ranking, we used the underlying TREC relevance judgement to evaluate both the Mean Average Precision (MAP) and normalized Discounted Cumulative Gain (nDCG@10) metrics. These were chosen as they are well regarded in IR research and are optimized when ranking documents according to the Probability Ranking Principle, as is our objective function. 

We generated rankings for each query over $T = 500$ time steps (or query instances) and evaluated MAP and nDCG@10 at each time step. We repeated this experiment 100 times and averaged out the effect of the stochastic user, and then repeated the experiment for each of the 50 queries. We averaged our results and repeated the experiment for different values of $\lambda$, the results of which can be seen in Table~\ref{table1:TREC}.

From the table we can see that the exploration parameter $\lambda$ is indeed essential for tuning the model as performance when $\lambda = 1$ is consistently poor. Conversely, we also see that in the myopic case ($\lambda = 0$) we obtain very good performance, indicating that our framework alone (without any multi-armed bandit based exploration) is able to learn document relevancy over time and generate rankings that give good MAP and nDCG@10 scores across all click models. As $\lambda$ varies from 1 we see a marked improvement that is often better than the myopic case, before converging to the myopic score as $\lambda \to 0$. This suggests that for larger values of $\lambda$, the model suffers from too much exploration and doesn't generate exploitative rankings, whereas there exists an optimal amount of exploration that produces better rankings than the greedy case. 

As such, for future experiments we set $\lambda = 0.1$ as this provided good results for each click model in both metrics. The nDCG@10 metrics showed that smaller values of $\lambda$ were optimal, but as the metric was less discriminative we gave greater focus to the MAP scores instead. 

\subsection{Click Log Experiment}

Following our encouraging simulation analysis, we experimented using a click log data set so that we could 1) use the click log to learn document relevancies in advance before generating rankings; 2) use realistic click model parameters learnt from the data; 3) use the log to evaluate the performance of our algorithm and provide an upper bound to compare against.

We use the Yandex Relevance Prediction Challenge data set\footnote{http://imat-relpred.yandex.ru/en/}, an anonymised search log containing 43,977,859 search sessions, each containing a query, the documents displayed at ranks 1 to 10 and any clicks on those documents. No additional document feature information is given, making it well suited to our problem definition of learning through user feedback alone. In addition, the log contained 71,930 relevance judgements for training purposes. For this experiment, we narrowed down the 30,717,251 unique queries to 1,327 queries that were searched for in over 1000 sessions and contained at least 10 relevance judgements. 

We performed a similar experiment to the TREC simulation, where we used different click model variants of the UCB-IE algorithm to generate rankings of $M = 10$ documents. We restricted our analysis to two values of $\lambda = 0$ and 0.1 (learnt from the simulation experiment) so as to evaluate the effect of exploration, and we used the known relevance judgements to evaluate MAP and nDCG@10 at each time step. 

In order to demonstrate and evaluate the ability for the algorithm to learn from existing click data, we repeated the experiment but introduced a training phase over the first 50\% of the data, whereby no rankings were generated but the ranking and clicks found in the data were interpreted by the model and used to update $\hat{r}_d$ and $\gamma_d$. 

\subsubsection{Interpreting Clickthroughs in the Data}
\begin{table}
\centering
\begin{tabular}{ c | c | c | c}
$i$ & UCB-IE before & Data set & UCB-IE after \\
\hline \hline
1 & A & \textbf{B*} & B \\
2 & B & E* & D \\
M = 3 & C & \textbf{D*} & E \\
\hline
$> M$ & D & & \\
 & E & & 
\end{tabular}
\caption{Example demonstrating the restriction of the data set. A $\to$ E are documents chosen by UCB-IE to display at rank positions $i$ at time $t$. Bold indicates that a click occurred for that document at that position in the dataset, and a star indicates the information that is used in the UCB-IE update.}
\label{table2:Example}
\end{table}

\begin{table*}[t]
 \centering
\begin{tabular}{c | c | c | c | c | c | c | c }
 \multicolumn{8}{c}{ \textbf{MAP 0\% training phase}} \\
   \hline  \hline
& Upper Bound & MC &  MC-Prior & EH & EH-Prior & DCM & DCM-Prior \\
\hline
$\lambda = 0.1$ & 0.7006 $\pm$ .056 & 0.5657 $\pm$ .061 & 0.5636 $\pm$ .064 & 0.5374  $\pm$ .062 & 0.5339 $\pm$ .064 &  0.6320 $\pm$ .061 & 0.6306 $\pm$ .061    \\
$\lambda = 0$ &  & 0.5998 $\pm$ .063 & 0.5997 $\pm$ .066 & 0.5713 $\pm$ .064 & 0.5698 $\pm$ .065 & \textbf{0.6368 $\pm$ .061} & 0.6354 $\pm$ .063\\
\hline
 \multicolumn{8}{c}{\textbf{50\% training phase}} \\
   \hline  \hline
$\lambda = 0.1$ & 0.7006 $\pm$ .056 & 0.6360 $\pm$ .044 & 0.6353 $\pm$ .045 & 0.6178  $\pm$ .046 & 0.6161 $\pm$ .046 & 0.6582 $\pm$ .046 & 0.6576 $\pm$ .047 \\   
$\lambda = 0$ &  & 0.6519 $\pm$ .045 & 0.6519 $\pm$ .046  &  0.6354 $\pm$ .046 & 0.6348 $\pm$ .047 & \textbf{0.6614 $\pm$ .047} & 0.6607 $\pm$ .047
\end{tabular}
\begin{tabular}{c | c | c | c | c | c | c | c }
 \multicolumn{8}{c}{\textbf{nDCG@10 0\% training phase}} \\
   \hline  \hline
& Upper Bound & MC &  MC-Prior & EH & EH-Prior & DCM & DCM-Prior \\
\hline
$\lambda = 0.1$ & 0.8334 $\pm$ .046 &  0.7248 $\pm$ .058 & 0.7244 $\pm$ .057  & 0.6905 $\pm$ .057 & 0.6900 $\pm$ .057 & 0.7780 $\pm$  .056 &  0.7772 $\pm$ .056    \\
$\lambda = 0$ &  & 0.7466 $\pm$ .057 & 0.7460 $\pm$  .057 & 0.7206 $\pm$ .056  & 0.7208 $\pm$ .056 &  \textbf{0.7825 $\pm$ .056} & 0.7818 $\pm$ .056 \\
\hline
 \multicolumn{8}{c}{\textbf{50\% training phase}} \\
   \hline  \hline
$\lambda = 0.1$ & 0.8334 $\pm$ .046 & 0.7248 $\pm$ .058 & 0.7244 $\pm$ .057 & 0.6905 $\pm$ .057   &  0.6900 $\pm$ .057 &  0.7758 $\pm$ .056 & 0.7753 $\pm$ .056  \\   
$\lambda = 0$ &  & 0.7466 $\pm$ .057 &  0.7460 $\pm$ .057   & 0.7206 $\pm$ .056 & 0.7208 $\pm$ .056 &  \textbf{0.7797 $\pm$ .056}  & 0.7788  $\pm$ .056
\end{tabular}
\caption{Mean final MAP and nDCG@10 values $\pm$ variance for each UCB-IE click model variant, for both the explorative ($\lambda = 0.1$) and myopic cases ($\lambda = 0$), and for different training regimes. Optimal values are in bold.}
\label{table3:clickLog}
\end{table*}

During the evaluation phase, our model needs to observe clickthroughs for its generated rankings so that it can update its probability of relevance estimate ready for the next ranking. In the previous experiment, we achieved this by simulating a user clicking on documents; in this experiment we used the clickthroughs contained in the data. Unfortunately, this presented a problem as it was often not the case that at time $t$ (or session $t$) the document ranking in the data set contained a particular document and subsequently a click event. Initial results showed very poor performance due to the fact that even when the UCB-IE was discovering relevant documents and displaying them correctly, if the data set hadn't displayed the document as well then no click event would occur and the document would be penalised by the model update and not shown again. As such, the algorithm's effectiveness was hampered by the rankings contained in the data set.
 
To counter this we limited the UCB-IE algorithm to only display documents that were already ranked in the data at time $t$, but ranked according to its own calculated probability of relevance. In this way, we were never faced with the issue of not being able to provide feedback for a ranked document. In addition, when updating the probability of relevance we interpret the clickthroughs and rank positions of the ranking in the dataset, not the ranking generated by UCB-IE. This is illustrated in Table~\ref{table2:Example}, where the third column represents the `restricted' ranking that would be generated, a re-ranking of that found in the data set.

This restricted interpretation leads to the following limitations: 1) We can only perform as well the dataset ranking, as we are not able to discover different, relevant documents and receive user feedback on them, thus stifling exploration. Thus, we consider the ranking in the data set as an upper bound; 2) Interpreting the clicks in the data rather than the UCB-IE's ranking may lead to a slow learning rate, for instance the case where a relevant document is ranked in a low position in the data but ranked highly by UCB-IE.
 
\subsubsection{Experiment and Analysis}

Before running each experiment, we used the query search log data and relevance judgements to estimate the parameters for each of the click models. The parameter values for $\pi_i$ and $b_i$ were calculated at each rank for each query using maximum likelihood estimators and simple counts (such as occurrences of clicks at rank $i$ for relevant documents etc.). In this way we do not make any prior assumptions on the click model and the query specific parameters are able to capture the different characteristics of each query, for example, the clicking behaviour for navigational queries is different to that of informational queries \cite{Lee:navigational}, although this simple technique does run the risk of overfitting. A more sophisticated and accurate method would be to use probit Bayesian inference \cite{Zhang:probit}, although for the purpose of this experiment this was not necessary.   

Another aspect of the model that we wanted to test was its responsiveness to the addition of new documents during its run time. Thus, for each click model we introduced two variants, the original algorithm which adds a new document whenever it first occurs in the data set, and a `Prior' version that has a full list of documents in advance. We summarise the results of our experiments in Table~\ref{table3:clickLog}:

The `Upper Bound' in the table refers to the ranking found in the data set. As previously mentioned due to the limitations of interpreting the data set clickthroughs we can not expect to exceed the upper bound performance, as well as the fact that we do not know the underlying mechanism used to rank the documents and that the model used had access to significantly more information such as document features. Our experiment is concerned with analysing the performance of the different click models, the parameter $\lambda$ and the ability to incorporate new documents. 

Some notable observations: 1) Generally the Examination Hypothesis model proved a bad fit for the data, whereas the Dependent Click model performed well; 2) The Dependent Click Model consistently had the best results, which were closest to the upper bound 3) The `prior' variants performed marginally worse than their equivalents, implying that the algorithm is able to respond to the arrival of new documents and learn their probability of relevance accordingly over time; 4) Introducing a training period greatly improved the scores and demonstrated how the model could be used to learn from existing click data in a practical application; 5) As expected, the limitations imposed on interpreting clicks stifled exploration and we find that the myopic case performed better than the explorative case. Ultimately, this experiment showed promising but inconclusive results, and serves to demonstrate the difficulty in evaluating an online algorithm that is responsive to user behaviour. 

\section{Discussions and Conclusions} 
\label{conclusions}
We have presented a probabilistic optimisation framework for multi-period information retrieval, where unlike offline IR techniques, here we have opportunities to learn the relevance of documents over a period of time through the interaction with users (by considering clicks as observations of relevance). Unlike previous studies, we consider the following assumption about the users' feedback: the probability of a viewer examining and clicking a document is dependent on its ranking position and can be described using a click model. Through the theoretical derivation and analysis, we showed how the belief about the relevancy of documents (the posterior probability of being clicked) evolves over time from the rank-biased user feedback (clicks). Besides the theoretical understandings and insights, three click model dependent, practical stochastic ranking strategies have been demonstrated on the basis of the bandit machine theory. 

Using TREC simulations we were able to analyse the strategies and assess the impact of exploration. This information informed our next experiment, where we attempted to demonstrate the performance capabilities of our model and some variants despite the restrictions imposed by the underlying data set. We also showed that the model could be used to learn from existing data to boost performance, and that it is robust to the discovery of new documents. 

We will continue to seek appropriate data that could be used to properly evaluate the algorithms performance, or the use of a search engine to experiment on, and we will also continue to determine ways in which to overcome the limitations of existing data. We could also use more sophisticated techniques to determine the click model parameters, or plug-in newer click models into the framework and analyse whether there is improved performance. 

We are also interested in using document features to provide context when learning over time, and comparing our algorithm with learning to rank algorithms. In addition, in order to decrease the number of time periods it takes to learn an optimal ranking we intend to use these features to explore how document similarity could be used to update the probability of relevance for non ranked documents that are similar to ranked documents. This could then be expanded to iteratively learn a diverse and relevant set of documents over time by making use of the portfolio theory of IR \cite{Wang09portfoliotheory}.

While helping us develop an insight to the optimal document ranking over time, the probabilistic framework has a number of strong assumptions that may not necessarily correspond with established observed user browsing practice, such as maximising the number of clicks.  Additionally, there are limitations on how many click models can be incorporated into our general mixed click model, but we are confident that most mainstream models (some of which we have demonstrated) can be included. Another drawback is that the parameters for the plugged-in click model have to be learned in advance and must fit the observations received by the algorithm in order to be effective, a non-trivial task particularly in the absence of relevance judgements. 

We could also expand our MAB research further, making use of Exp3 \cite{exp3} or a restless bandit (bandits whose distributions change over time) formulation \cite{restlessbandits}, or one of the improved UCB variants \cite{ImprovedUCB, regretBounds}, and then conduct an experiment with changing document relevancies. We are also working on an optimal dynamic programming solution that can be approximated using a partially observable Markov decision process, which should give a firmer theoretical justification for the optimality of the algorithm. 

\vspace{1 mm}
\begin{scriptsize}
\bibliographystyle{acm}
\bibliography{Refs}  
\end{scriptsize}

\balancecolumns
\end{document}